# The Challenging History of Other Earths[*]

Christopher M. Graney
Vatican Observatory
00120 Stato Città del Vaticano
c.graney@vaticanobservatory.org

**Abstract**

**This paper provides an overview of recent historical research regarding scientifically-informed challenges to the idea that the stars are other suns orbited by other inhabited earths—an idea that came to be known as "the Plurality of Worlds". Johannes Kepler in the seventeenth century, Jacques Cassini in the eighteenth, and William Whewell in the nineteenth each argued against "pluralism" based on what in their respective times was solid science. Nevertheless, pluralism remained popular despite these and other scientific challenges. This history will be of interest to the astronomical community so that it is better positioned to avoid difficulties should the historical trajectory of pluralism continue, especially as it persists in the popular imagination.**

**Introduction**

The idea that the universe contains an abundance of worlds like Earth, home to intelligent life, has historically been challenged by science. This matters for the modern astronomical community, because popular imagination strongly embraces this idea, thanks both to its history and its ongoing promotion by astronomers themselves. This paper provides an overview of recent historical research regarding the idea of other earths, focusing on how Johannes Kepler (1571-1630), Jacques Cassini (1677-1756) and William Whewell (1794-1866) each used science to criticize this idea.

The idea that other earths might exist arguably begins with Nicolaus Copernicus (1473-1543) and Giordano Bruno (1548-1600): Copernicus envisioned Earth circling the sun along with Jupiter, Mars, etc.; Bruno, a Copernican, envisioned stars being other suns, circled by other inhabited earths (Crowe, 2008: 37-38, 46-47). Modern science considers the possibility of detecting such inhabitants, through SETI, for example. This draws the attention of the broader public, as the idea of other earths has been broadly popular since Bruno. However, the work of Kepler, Cassini, Whewell and others illustrates how historically that popularity seems to have

---

been driven by politics or even religion more than by scientific discoveries. Therefore, their work offers a caution to astronomers who themselves are often drawn to the popular idea of other earths and intelligent life on them.

**Stars that are not Suns**

Johannes Kepler embraced the idea of extraterrestrial intelligent life. He wrote favorably about the possibility of intelligent life on Jupiter, for example, supposing that all its moons surely served the Jovians well, illuminating the skies of that stupendous world (Kepler, 1965: 39-47). But he would have rejected SETI on scientific grounds. He determined that there was but one sun in the universe—ours.

Kepler said in his 1610 *Conversation with Galileo's Starry Messenger* that the simplest observations, measurements, and calculations proved Bruno wrong. "To use Bruno's terms, the [fixed stars] are suns," Kepler wrote, "Nevertheless, let him not lead us on to his belief in infinite worlds, as numerous as the fixed stars and all similar to our own." Noting that Galileo reported over 10,000 telescopically visible stars, Kepler continued (Kepler, 1965: 34-35),

> *The more there are, and the more crowded they are, the stronger becomes my argument against the infinity of the universe.... Suppose that we took only 1,000 fixed stars, none of them larger than 1´ (yet the majority in the catalogues are larger). If these were all merged in a single round surface, they would equal (and even surpass) the [apparent] diameter of the sun. If the little disks of 10,000 stars are fused into one, how much more will their visible size exceed the apparent disk of the sun?*

Astronomers from Ptolemy in the second century to Tycho Brahe in the sixteenth had repeatedly measured the apparent diameters of visible stars to be ~1 arc minute (Van Helden, 1985: 15-53), unaware that these diameters were optical artefacts and thus did not reflect physical sizes (unlike the apparent diameters of the sun and moon). Kepler continues:

> *If this is true, and if they are suns having the same nature as our sun, why do not these suns collectively outdistance our sun in brilliance? Why do they all together transmit so dim a light…? When sunlight bursts into a sealed room through a hole made with a tiny pin point, it outshines the fixed stars at once. The difference is practically infinite.*

In other words, Kepler argued that stars are too dim to be bodies like our sun. Their light is infinitely weaker, practically speaking, despite their combined apparent size rivalling the sun. And since astronomers understood the stars to have measurable apparent diameters, the distances of the stars could not explain this weak light. "Will my opponent tell me that the stars are very far away from us?" Kepler wrote (Kepler, 1965: 35), "This does not help his cause at all. For the greater their distance, the more does every single one of them outstrip the sun in diameter." Stars had to be huge, given their measured apparent sizes and their vast distances from Earth as required by the Copernican system (which Kepler strongly supported) absent detectable stellar parallax. In 1604, Kepler had calculated that Sirius surpassed the orbit of Saturn in size. Every single star visible to the eye had to surpass the orbit of Earth (Figure 1)—and be far dimmer than the sun (Kepler, 2021; Graney, 2019).

The stars simply were not suns (Koyré, 1957: 63-86): "Our sun," Kepler wrote in the *Conversation* (Kepler, 1965: 43), "is more splendid than the fixed stars." Kepler made it clear that Bruno's ideas had no basis in science. Every astronomer could make basic measurements, do simple math, and reproduce Kepler's results to see that Bruno was wrong and there was only one sun (Graney, 2019).

**A Plurality of Worlds**

The universe we envision today is closer to Bruno's than to Kepler's. We search for intelligent life on other earths orbiting other suns, not on Jupiter. Bernard Le Bovier de Fontenelle promoted the idea of other earths with his 1686 book, *Conversations on the Plurality of Worlds*. Its frontispiece, by Juan d'Olivar, showed the solar system as but one system among many (Figure 2). "French society was shocked," says historian Lucía Ayala (Ayala, 2012), and "this shock quickly spread throughout Europe."

*Plurality* discussed how the moon must be inhabited because it is like the Earth, and the planets, too, because they are like the moon. Invoking the theory of René Descartes, Kepler's younger contemporary, in which vortices in a universal fluid drove orbital motion (Charalampous, 2019), *Plurality* supposed the stars, "luminous bodies in themselves, and so many suns", to be vortex centers with their own planets. These hosted inhabitants who see the universe just as we do (Fontenelle, 1715: 134-141). The universe comprised a "plurality of worlds" like ours.

Fontenelle's reasoning regarding stars being "so many suns" was that "their light is bright and shining"—and twinkling, because of interactions between stellar vortices (Fontenelle, 1715: 134, 150). He was apparently unaware that astronomers knew that twinkling is an atmospheric phenomenon; Christoph Scheiner and his student Johann Georg Locher had discussed it thoroughly, seven decades earlier (Scheiner & Locher, 2017). *Plurality's* science here was not rigorous.

However, Ayala argues that *Plurality*'s appeal was political, not scientific. King Louis XIV of France had embraced Copernicanism as a metaphor for his rule as "The Sun King". Ayala writes that he "undertook many endeavours to convince others of his own conviction: that natural forces legitimized his power, Copernicanism being his major proof." She references a flattering preface, written to him in an astronomy book: "Indeed, SIRE, you are at the center of this kingdom as the sun, according to the hypothesis of Copernicus, is at the center of the *universe*", etc. (italics added). This "colonization" of the sun by Louis XIV, Ayala argues, "paved the way for the enthusiastic reception of the idea of the plurality of worlds", dethroning the sun and diminishing the Sun King. D'Olivar's illustration features a face in every one of the innumerable suns (Figure 3). Our sun is not special; the universe is not structured around a single, brilliant ruler (Ayala, 2015: 207, 209, 216).

Politics aside, by 1686 the scientific case for Kepler's single, brilliant sun happened to be weakening. Telescopes had once seemingly confirmed the single-sun universe that appealed to Louis XIV. In the first half of the seventeenth century, various astronomers, including Galileo, Scheiner and Locher, Simon Marius, and Giovanni Battista Riccioli, observed stars telescopically (with apertures of 1-2 cm, typically), and recorded them as disks of measurable apparent sizes (Figure 4). Astronomers were again unaware that these disks were optical artefacts, in fact the Airy disks produced by such small apertures owing to the wave nature of light, a phenomenon unknown at that time.

The disk sizes, ~10 arc seconds for the more prominent stars, were substantially smaller than pre-telescopic measurements, but still translated into giant stars in a Copernican universe, especially since telescopes increased sensitivity to parallax. Galileo supported Copernicus, of course, but struggled with the giant star question. Scheiner, Marius, and Riccioli all argued for Brahe's Earth-centered model of the universe that accommodated new telescopic discoveries while having stars located just beyond, and thus like in size to, Saturn (Graney, 2015: 45-86, 129-139; Graney, 2021).

But by the 1660s, Johannes Hevelius and Christiaan Huygens had published observations (of lunar occultations of stars, and of stellar apparent sizes diminishing with filtering) suggesting that even telescopes greatly inflated stellar sizes. In such a case, stars could be suns. These observations did not go unchallenged, however. For example, John Flamsteed, England's first Astronomer Royal, countered that Sirius and Mercury, seen through the same small aperture telescope under the same conditions, had the same apparent diameters (Graney, 2015: 148-158; Applebaum, 2012). Thus when Fontenelle published *Plurality*, the science concerning stars being suns was in flux.

But in 1716 and 1740, Jacques Cassini of the Paris Observatory published high-quality telescopic (small aperture) measurements of Sirius's apparent diameter: 5″. He showed Sirius to be giant, granted a Copernican universe. As with Kepler, all stars had to be giant—not suns (Graney, 2021).

Fellow astronomers acknowledged Cassini's measurements, but enthusiasm for a plurality of worlds, or "Pluralism", had caught on—they considered stars to be suns anyway. John Hill borrowed heavily from Cassini in his 1754 book *Urania*, and wrote how—
> *The observation of Sirius's diameter being five seconds had, for its author, one of the most accurate, and most judicious astronomers the world has ever known, Cassini, and, whenever it is repeated with the same apparatus, it succeeds in the same manner, and verified very punctually; and other stars have also apparent diameters of nearly the same extent.*

Note the emphasis on the reproducibility of Cassini's results. Nevertheless, Hill wrote that each star is a sun, with "earthy planets rolling round it"—because, why else would the stars exist, and doesn't such a universe glorify its Creator? Pluralism apparently had religious appeal for Hill (Graney, 2021; Hill, 1754). Whether Pluralism's appeal was political or religious, it was such that high-quality, easily-reproducible telescopic data, even from the esteemed Cassini, would not overcome it.

Of course, in time it would become clear that Huygens was right about star sizes, and Cassini, Flamsteed, and Kepler were wrong. Full understanding that telescopic stellar disks were optical artifacts, not stellar bodies, had to await the nineteenth century, the wave theory of light, and George Biddell Airy (Graney, 2015: 154-157; Graney & Grayson, 2011).

By then Pluralism was broadly accepted; astronomers and others in the early nineteenth century typically considered every planet in orbit around the sun to be inhabited. William Herschel, the first astronomer to discover a new planet, had even proposed that the sun was inhabited, "like the rest of the planets". "Astronomical principles" showed this, he wrote in 1795, and he was sure he could answer every objection made against his proposal. Thomas Paine in his *Age of Reason* wrote of the inhabitants of "each of the worlds of which our [solar] system is composed" and of those "millions of worlds" around other stars, also. In the 1830s Thomas Dick even estimated the population of the solar system: tens of trillions (not counting solar inhabitants). The population of the universe? Trillions of times larger still (Crowe, 2008: 179-181, 228-229, 269-271; Crowe, 1986).

**A Diversity of Stars**

Once the star size problem was resolved, science ceased to directly challenge Pluralism, but science offered nothing to support it, either. Toward the end of the nineteenth century, Agnes Clerke described the progress of knowledge about stars during that century. At the century's start, little was known; stars were a background against which astronomers measured planetary motions. "The sidereal world... was the domain of far-reaching speculations" that were unencumbered by systematic study. William Herschel, she noted, assumed that "the brightness of a star afforded an approximate measure of its distance"—one star was more or less the same as another (Clerke, 1908: 11, 21). This meshes with stars being (more or less identical) suns. With so little known about the stars, nothing stood against this assumption, but no science supported it, either.

Yet by the century's end, Clerke wrote, the distances to roughly one hundred stars were known through parallax measurements. Moreover (Clerke, 1908: 37):

> *the list [of stars with measured distances] is an instructive one, in its omissions no less than in its contents. It includes stars of many degrees of brightness, from Sirius down to a nameless telescopic star in the Great Bear.*

Many of the brightest stars had been found to be too remote to measure their distances, while many stars that were found to be nearby were quite faint:

> *The obvious conclusions follow that the range of variety in the sidereal system is enormously greater than had been supposed, and that estimates of distance based upon apparent magnitude must be wholly futile. Thus, the splendid Canopus, Betelgeux, and Rigel can be inferred, from their indefinite remoteness, to exceed our sun thousands of times in size and lustre; while many inconspicuous objects, which prove to be in our relative vicinity, must be notably his inferiors. The limits of real stellar magnitude are then set very widely apart.*

We have learned much since Clerke. Still, her observation holds true. There is great diversity in the universe of stars, from those rare stars that rival the sizes Kepler determined, down to the super-abundant red dwarfs.

### A Diversity of Planets

The nineteenth century revealed a great diversity among planets, too. The historian of science Michael J. Crowe has argued that William Whewell, Master of Trinity College of Cambridge University, first envisioned this diversity. In 1853, Whewell anonymously published *Of the Plurality of Worlds: An Essay*, attacking scientifically the idea of a universe widely inhabited with intelligent life (Crowe, 2016).

One of Whewell's key arguments was the inverse-square law; the heat and light a planet receives depends strongly on its distance from the sun. Crowe argues that Whewell was among the first to consider the idea of a "habitable zone" around a star (Whewell used the term "temperate zone"). This implied that the rest of the region around a star is *not* habitable, contrary to William Herschel, Thomas Dick, etc.

Whewell, formerly president of England's Geological Society, also used geology to argue against widespread extraterrestrial intelligence. Crowe writes (Crowe, 2016: 441):

> *Whewell's argument was that evidence for the age of the Earth showed that throughout most of Earth's history it had been bereft of intelligent life, which suggested that the Creator's plan for the cosmos was capacious enough to leave vast regions of it lacking [intelligent extraterrestrials] for long periods of time.*

Whewell speculated nevertheless about lower life within the solar system—such as, on Jupiter, "aqueous, gelatinous creatures; too sluggish, almost, to be deemed alive, floating on their ice-cold water, shrouded forever by their humid skies (Whewell, 1853: 185-186)."

Scientific evidence for diverse planets, and not merely other earths, had long been available. The great diversity in planetary sizes had been known for more than two centuries (Figure 5). Isaac Newton, whose physics supplanted Descartes's vortices, had included in the third edition of his *Principia* (1726) calculations showing substantial differences among solar system bodies in mass, density, and surface gravity (Figure 6); differences from Earth might have provided reason to suppose that the planets would lack intelligent life (Crowe, 2016).

Astronomers before Whewell had recognized planetary diversity, but in ways inclined toward Pluralism. Consider the prominent astronomer John Herschel, Whewell's friend and William's son. Prior to Whewell's 1853 *Essay*, Herschel had discussed Mercury receiving more solar heat than Uranus, Jupiter having stronger gravity than Earth, and the density of Saturn being so low that it "must consist of materials not much heavier than cork". Yet he presumed that the planets would be inhabited, much as his father presumed the sun would be inhabited; he marveled at the "Benevolence and Wisdom which presides over all" the habitable universe, much like Hill (Crowe, 2016: 440). Writing to Whewell in response to the *Essay*, John Herschel stated that though he had compared Saturn to cork, "it *never did* occur to me to draw the conclusion that *ergo* the *surface* of Saturn must be of extreme tenuity." He then insisted that, even were Jupiter an ice-cold sea, it might be populated by something more than Whewell's gelatinous creatures; intelligent fish might construct "crystal palaces" (referencing the Crystal Palace of the London Great Exhibition of 1851) in the warmer depths of the Jovian seas (Crowe, 2008: 358-359).

But this inclination toward Pluralism was running into more scientific difficulties in the nineteenth century than just the inverse-square law and geological timescales. The idea that life spontaneously generated from matter, which would make Jovian fish or even life on the sun reasonable, was being overthrown.

This idea dated back to at least Aristotle. Frogs were thought to be generated right out of the mud (Deichmann, 2012); an ancient poet wrote of how "feetlesse and without legs on earth they lie (Topsell, 1658: 719-720)". Ancient Jewish Rabbis discussed whether the earth from which a mouse might be spontaneously forming would be unclean, since Leviticus 11:29 lists mice as being among the various "creeping things" that are unclean (van der Horst, 1993/94). Kepler remarked on how the Earth "kindles daily so many little living things from herself—plants, fishes, insects (Kepler, 2021: 58)"—and he apparently supposed that Jupiter kindled them, too. As late as the 1880s, the botanist Carl Nägeli argued for the spontaneous generation of the simplest living organisms, with higher life forms evolving from those (Farley, 1972; Mason, 1962: 428-31). But by the end of the nineteenth century science had rejected spontaneous generation entirely (Deichmann, 2012). Indeed, science today is very far from spontaneous generation, theorizing that life on Earth originated long ago, through a process neither reproducible in a lab nor found re-occurring in nature, and that more complex life originated in a merger of two primitive, single-celled organisms—a merger that occurred once in Earth's history, and only after primitive life had existed for many hundreds of millions of years (Lane, 2015).

So we do not expect to find life on the sun or Jupiter. Even one small colony of microscopic gelatinous creatures in the dark, ice-cold waters of Enceladus would be a monumental find. Solar system bodies differ from Earth. Exoplanets show us further diversity, with planets larger than Jupiter in orbits smaller than Mercury's, and with planets of sizes unlike any found in our system. Whewell was right about planetary diversity, beyond his own expectations.

**Conclusion**

History shows that we are drawn to envision other earths by some factors other than science. The idea of inhabited earths orbiting myriad other suns became popular while observations, and astronomers like Kepler and J. Cassini, stood against it. As later observations showed diversity among stars and planets, and showed life to not be an ordinary product of matter, enthusiasm for a "plurality of [Earth-like and inhabited] worlds" persisted. "Pluralism" has always been challenged by science. Indeed, it is important to remember our current understanding of the plurality of worlds: the sun is indeed a star, but few stars are like our sun; the Earth is indeed a planet, but among the many planets discovered so far, none is really like our Earth.

Today's astronomical community should be aware of this history of science, given that the discovery of "Earth-like planets" has become almost a cliché of astronomy department press releases, and some astronomers have proposed extraterrestrial intelligence in mainstream scientific publications to explain the anomalous behavior of objects, with the news media paying close attention (Editors, 2018; Clery, 2020; Wright & Gelino, 2018; Wright, et al., 2016; Loeb, 2021; Sheikh, et al., 2021; Ivanov, et al., 2020). History cautions us here. Enthusiasm for extraterrestrial intelligence as an explanation for "canal" features supposedly visible on Mars in the late nineteenth century became an embarrassment for astronomy and arguably had a long-term negative impact on planetary science (Doel, 1996: 13-15; Sagan & Shklovskiĭ, 1966: 276; Crowe, 1986: 543-546).

The argument of this paper is *not* against the hunt for possible inhabitants of other earths. It is certainly not against the hunt for extraterrestrial life in general, such as simple microbial life, as though history somehow urges that science overall is opposed to the existence of other earths and life on them. The argument of this paper is *for* awareness of the challenging history of the idea behind the hunt. In that way, if the historical trajectory of this idea continues, and we continue to find that the universe is not as we have always tended to envision it to be, we will be prepared to address that result—and to address a surely disappointed public, whose imagination has been stoked by centuries of enthusiasm for pluralism continuing through today—without the embarrassment and negative impact that comes from having publicly pursued an idea in ignorance of its, and our, long history.

**Declaration of Interest statement**

The author declares that he has no known competing financial interests or relationships with third parties whose interests may be affected by the content of the manuscript that could have appeared to influence the work reported in this paper.

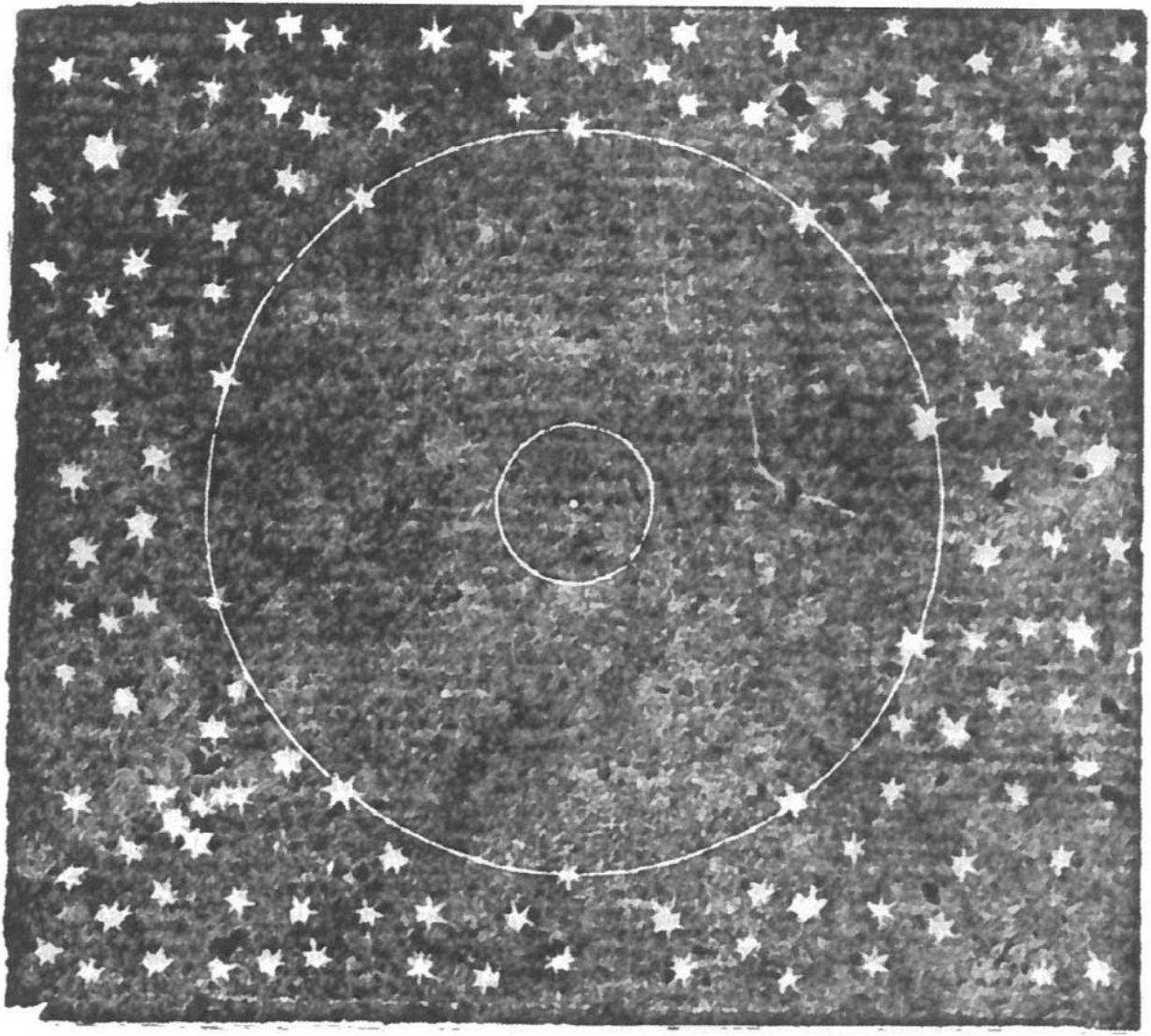

FIGURE 1: Diagram of the universe of stars from Johannes Kepler's *Epitome of Copernican Astronomy*, showing a small sun (the dot at the center) surrounded by large stars.

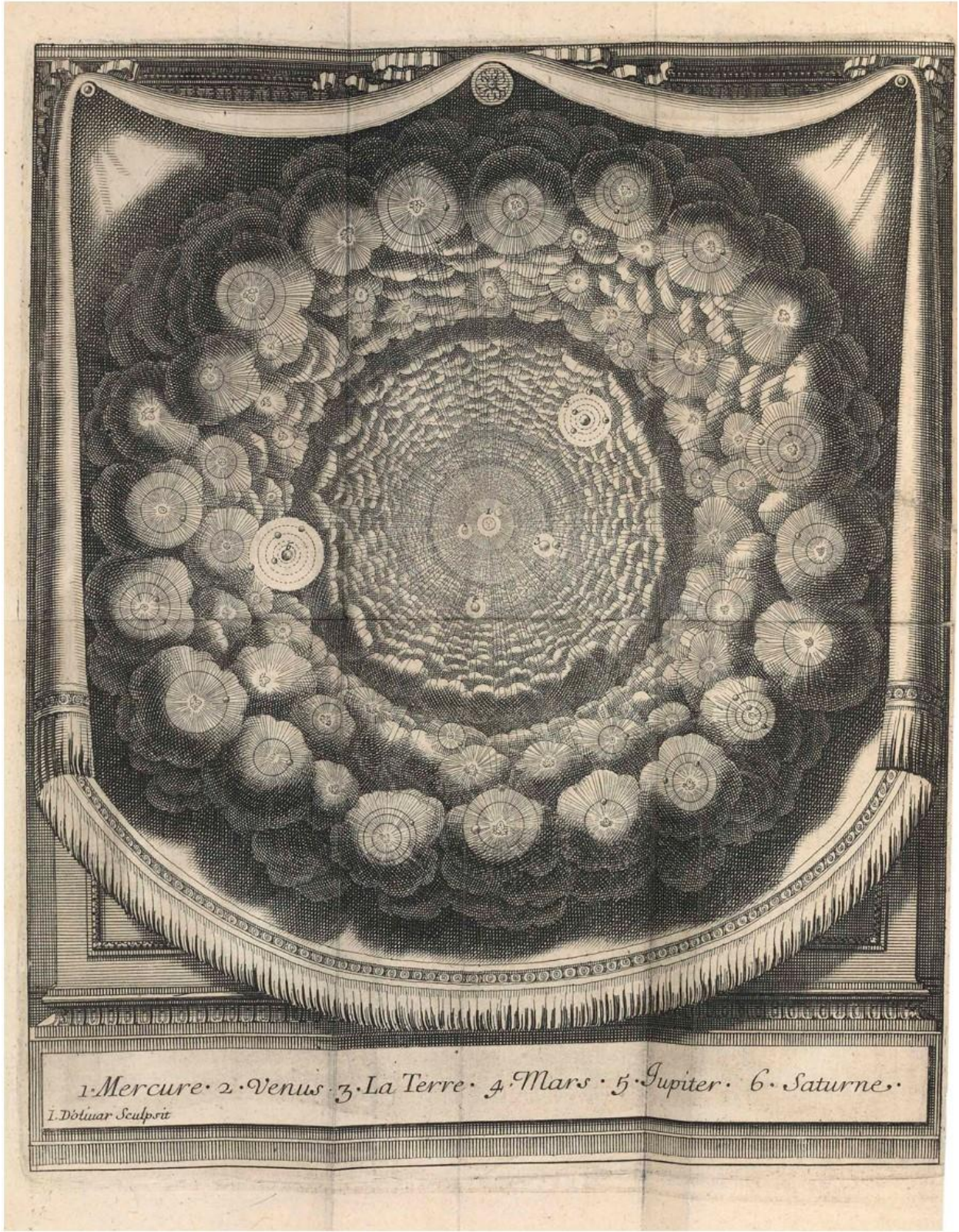

FIGURE 2: Frontispiece to Fontenelle's 1686 *Conversations on the Plurality of Worlds*.

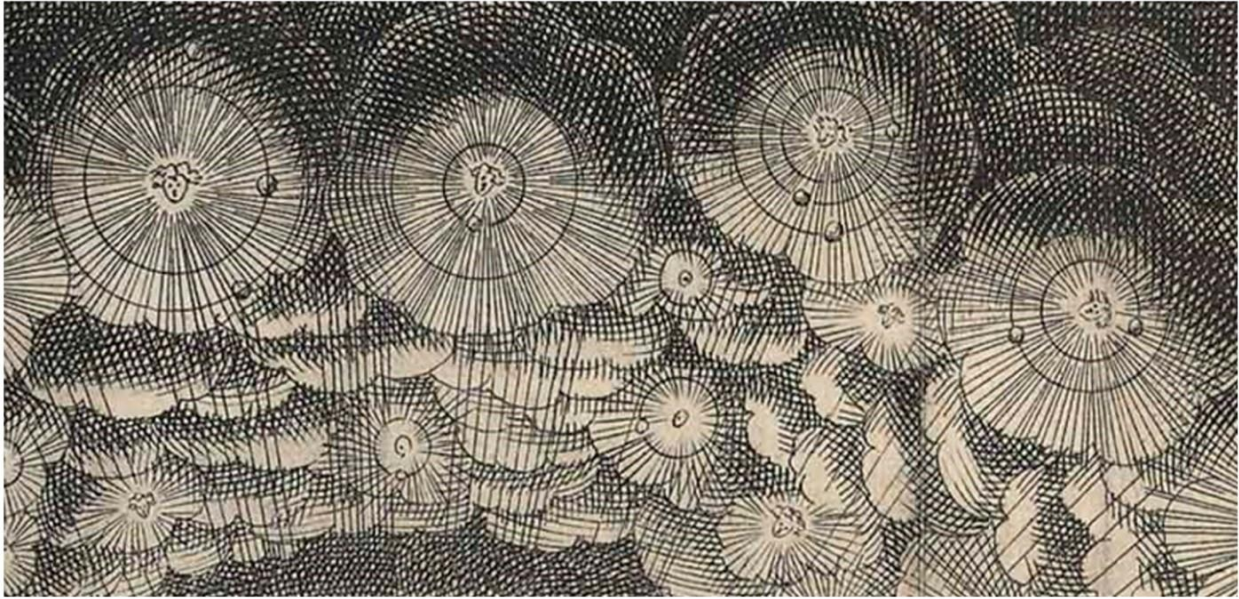

FIGURE 3: Detail from the Fontenelle frontispiece.

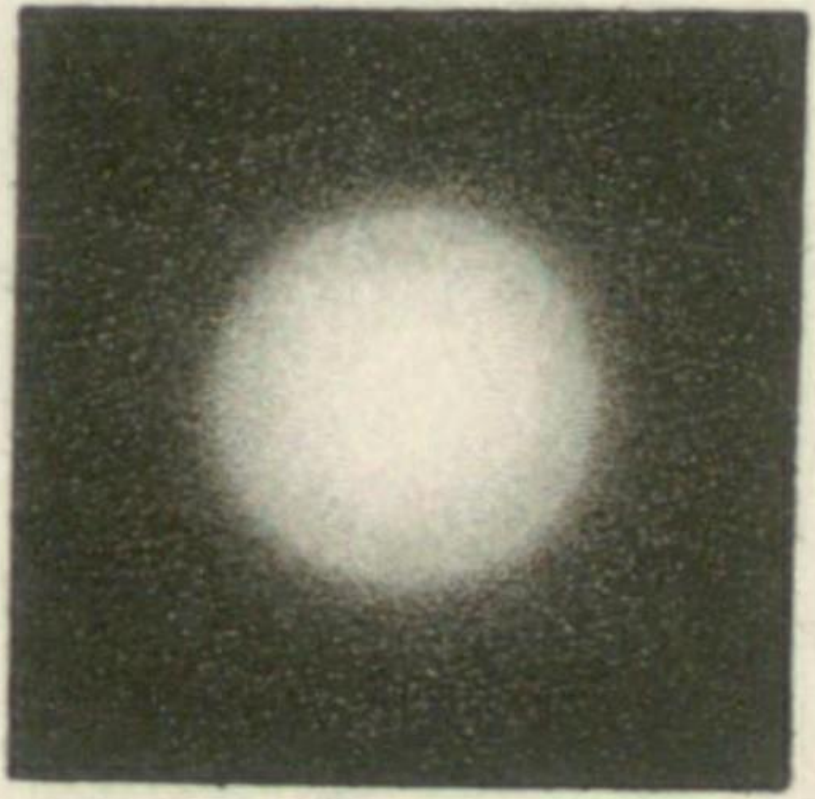

FIGURE 4: Illustration of a star as seen through a small-aperture telescope, from John Hershel's *Treatises on Physical Astronomy, Light and Sound*.

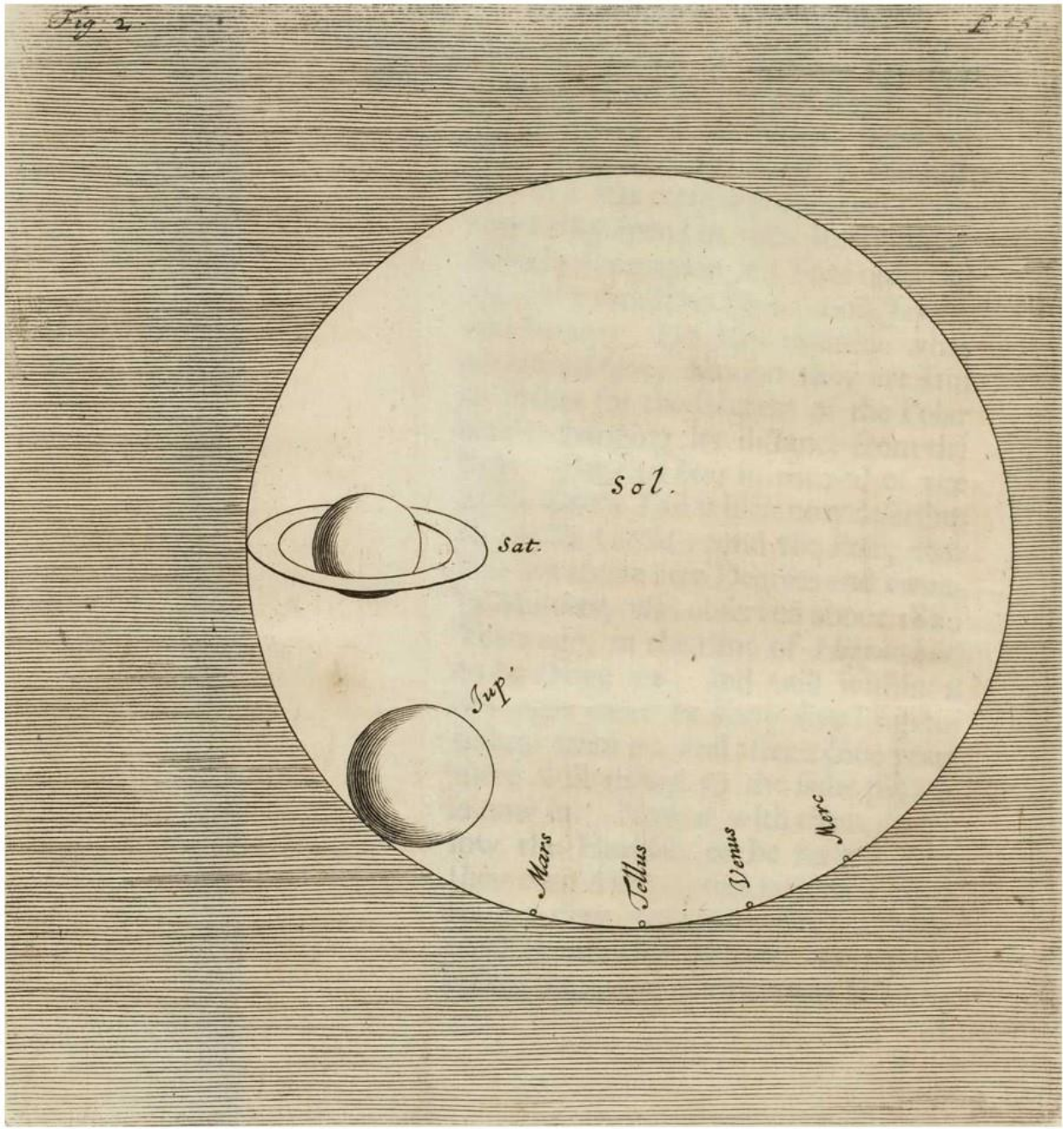

FIGURE 5: Sizes of planets and sun, from the 1698 *Cosmotheoros* of Christiaan Huygens.

genian satellite from the centre of Saturn, 3'. 4", and of the Moon from the Earth, 10'. 33"; by computation I found, that the weight of equal bodies, at equal distances from the centres of the Sun, of Jupiter, of Saturn, and of the Earth, towards the Sun, Jupiter, Saturn, and the Earth, were one to another, as 1, $\frac{1}{1067}$, $\frac{1}{3021}$, and $\frac{1}{169282}$ respectively. Then because as the

FIGURE 6: Strength of gravity on different bodies, from the 1729 edition of Newton's *Mathematical Principles of Natural Philosophy*, Volume 2 (p. 227).